\title {Quantum Computers: Noise Propagation and Adversarial Noise Models}
\author {  Gil Kalai\footnote{ 
Research supported in part by an NSF grant, 
an ISF grant, and a BSF grant. }
 \\
Hebrew University of Jerusalem and Yale University}
\documentstyle[12pt,amsfonts]{article}
\newtheorem{theo}{Theorem}

\newtheorem{lemma}[theo]{Lemma}
\newtheorem{prop}[theo]{Proposition}

\newcommand{\beq}[1]{\begin{equation}\label{#1}}
\newcommand{\enq}[0]{\end{equation}}

\begin{document}
\maketitle
\begin {abstract}

In this paper we consider adversarial noise models that will 
fail quantum error correction and fault-tolerant quantum computation. 

We describe known results regarding high-rate noise, sequential 
computation, and reversible noisy computation. We continue by discussing 
highly correlated noise and the ``boundary,'' in terms of correlation 
of errors, of the ``threshold theorem.'' Next, we
draw a picture of adversarial  
forms of noise called (collectively) ``detrimental noise.'' 

Detrimental noise is modeled after familiar 
properties of noise propagation. However, it can 
have various causes. We start by pointing out the 
difference between detrimental noise and 
standard noise models for two qubits 
and proceed to a discussion of highly entangled states, the rate of noise, 
and general noisy quantum systems.

\bigskip

\bigskip

\end {abstract}

\section {Introduction}
\label {s:in}

The feasibility of computationally superior quantum computers is one
of the most fascinating scientific problems of our time. 
The main 
concern regarding quantum-computer feasibility is that 
quantum systems are inherently noisy. 
The theory
of quantum error correction and fault-tolerant quantum computation (FTQC)
provides strong support for the possibility of building quantum
computers. In this paper we will discuss 
adversarial noise models that may fail quantum computation. 

This paper presents a 
critique of quantum error correction and skepticism on the 
feasibility of quantum computers. 
An early 
critique regarding noise and quantum 
computation (put forward in
the mid-90s by 
Landauer \cite {lan,lan2}, Unruh \cite {unr}, and others) 
was instrumental to the development of FTQC. 
Some of the ideas in the paper are 
provocative and speculative and they certainly {\bf do not} express 
established scientific material. We will also make some deviations 
from standard notation regarding quantum operations. We will use 
ordinary function notation for quantum operations 
(superoperators). So when $E$ is a quantum operation 
and $\rho$ is a state (described in terms of a density matrix), we 
will denote by $E(\rho)$ the action of $E$ on $\rho$. 

The paper describes in part my research, and relies on 
three (closely related) 
discussion papers \cite {DD,K2,K1}.
It has also benefited from several weblog discussions. 
Many colleagues
contributed helpful comments,
and allow me to single out Greg
Kuperberg for his ongoing, patient adversarial partnership, and Daniel Lidar 
for a careful editing of an earlier version of the paper.

We will now describe the structure of the paper. 
Section \ref {s:ex}  presents an example that can 
be regarded as a ``role model'' for adversarial noise.
Section \ref {s:bm} describes the basic framework for noisy 
quantum computers and the ``threshold theorem.''
In Section \ref {s:ob} 
we describe several obstructions in terms of the noise to 
quantum computation. We discuss high-rate noise, reversible 
noisy computation, and sequential noisy computation. 
In Section \ref {s:hc} we discuss
highly correlated noise and we also come back to the rate of noise. 
We try to draw a line between the 
types of correlations to which the ``threshold theorem'' applies 
and those to which it cannot apply.  
  
In Sections \ref{s:np}-\ref {s:gs}  
we propose a hypothetical form of
noise that we call ``detrimental noise'' 
that can cause quantum
error correction and fault-tolerant quantum computing to fail. 
Detrimental noise
is described via some (counterintuitive) properties. 
Perhaps the simplest way to think about our proposed detrimental 
noise picture is to regard the characteristics 
of noise propagation as the fundamental properties 
of noise in quantum systems, and to note 
that these properties can have other causes. 
We discuss noise propagation in Section \ref {s:np}.    

We describe detrimental noise starting with the case of two qubits (Section 
\ref {s:dn}), consider 
error synchronization for systems with many highly
entangled qubits, and  
discuss general open
quantum systems (Section \ref {s:gs}).

The conjectures toward the end of the paper can be regarded as proposed 
properties for noise models for quantum computers (and
more general quantum systems) that will cause 
quantum error correction and FTQC to fail.  Alternatively, the 
conjectures can be regarded as {\bf consequences} of 
a lack of fault tolerance in quantum 
systems. As such, they can be relevant 
to the nature of decoherence of 
quantum physical systems in nature even if computationally superior 
quantum computers are possible.

In Section \ref {s:cr} we discuss some concerns regarding our conjectures 
and, in particular, what their cause can be, and we return to the issue of 
modeling the rate of noise. In Section \ref {s:di} we  
discuss some further conceptual issues regarding noise and quantum computing.

\section {Example first!}

\label {s:ex}

Consider a quantum memory with $n$ qubits whose intended state is $\rho_0$.
Suppose that $\rho_0$
is a tensor product state. The noise affecting
the memory 
can be described by a quantum
operation $E_0$. Let us suppose that $E_0$ acts independently on
different qubits and its action on the $k$th qubit is as 
follows: with some small
probability $p$ the noise changes the state of the qubit into 
the 
completely mixed state $\tau_k$.

This depolarizing noise is a very simple form of 
noise that can be regarded as 
basic to the understanding of the standard models of noise as well as of
detrimental noise.

In the standard model of noise $E_0$ describes the noise of the
quantum memory regardless of the state $\rho$ stored in the memory.
This is quite a natural and indeed expected form of noise.

A detrimental noise will 
correspond to a scenario that, when the
quantum memory is supposed to be 
in a state $\rho$ and $\rho= U \rho_0$, the noise
$E$ will be $U E_0 U^{-1}$. Such noise is the effect of first
applying $E_0$ to $\rho_0$ and then applying $U$ to the outcome
noiselessly.

In reality we cannot perform $U$ instantly and
noiselessly and the most we can hope for is that $\rho$ will be the
result of a process. Our main conjecture  
is that for a noisy 
process intended
to lead to $\rho = U \rho_0$ the noise will contain a component of the form 
$E=UE_0U^{-1}$. 

Two remarks are in order: 1) The noise described by the quantum operation $E$ 
depends on the evolution of the 
quantum computer leading to $\rho$. The dependence of $E$ on the 
prior evolution is linear and there is nothing in this description that 
violates quantum mechanics linearity. In fact, this noise is a simple 
and familiar 
expression of noise propagation.
The quantum computer whose intended state is $\rho$ can 
be subject to a whole {\bf envelope} $D(\rho)$ of possible 
noise operations depending on the evolution leading to $\rho$. 
The relation between $D(\rho)$ and $\rho$ is nonlinear.

2) How can we claim that this 
example is damaging while noise propagation is successfully dealt 
with by fault-tolerant methods? We will 
discuss this question later.

\section {Quantum computers, noise, fault tolerance, 
and the threshold theorem}
\label {s:bm}

\subsection {Quantum computers}

The state of a digital
computer having $n$ bits is a string of length $n$ of zeros and ones.
As a first step toward quantum computers we can consider 
(abstractly) stochastic versions of digital computers where the state is 
a (classical) probability distribution on all such strings. 
Quantum computers are similar to these (hypothetical) stochastic 
classical computers and they work on qubits (say, $n$ of them). The state of 
a single qubit $q$ is described by a unit vector $u = a|0>+b|1>$  in 
a two-dimensional complex space $U_q$. 
(The symbols $|0>$ and $|1>$ can be thought of as 
representing two elements of a basis in $U_q$.) 
We can think of the qubit $q$ as representing 
$`0'$ with probability $|a|^2$ and $`1'$ with probability $|b|^2$. 
The state of the entire computer is a unit vector in the $2^n$-dimensional 
tensor product of these vector spaces $U_q$'s for the 
individual qubits. The state of the computer thus 
represents a probability distribution on the $2^n$ strings
of length $n$ of zeros and ones. The evolution of the quantum 
computer is via ``gates.'' Each gate $g$ 
operates  
on $k$ qubits, and we can assume $k \le 2$. 
Every such gate represents a 
unitary operator on $U_g$, namely 
the ($2^k$-dimensional) tensor product 
of the spaces that correspond to these $k$ qubits. At every ``cycle time''
a large number of gates acting on disjoint sets of qubits operate.

Moving from a qubit $q$ to the probability distribution on $`0'$ and $`1'$
that it represents is called a ``measurement'' and
it can be considered as an additional 1-qubit gate. We will assume
that measurement of qubits that amount to a sampling of 0-1 strings 
according to the distribution that these qubits represent is the final step 
of the computation.

\subsection {Noisy quantum computers}

The postulate of noise asserting that quantum systems are inherently noisy 
is essentially a hypothesis about
approximations. The state of a quantum computer can be
prescribed only up to a certain error. 
For FTQC there
is an important additional assumption on the noise, namely, on the
nature of this approximation.  The assumption is that the noise is
``local.'' This condition asserts that the way in which the 
state of the computer changes between 
computer steps is approximately 
statistically independent for different qubits. 
We will refer to such changes as ``storage errors'' or ``qubit errors.'' 
In addition, the gates that carry the computation itself are imperfect.
We can suppose that every such gate involves a small number 
of qubits and that the gate's
imperfection 
can take an arbitrary form, 
and hence the errors (referred to as ``gate errors'') created 
on the few qubits 
involved in a gate can be statistically dependent. 
We will denote as ``fresh errors'' to the storage errors 
and gate errors in one computer cycle.
Of course, qubit errors and gate errors propagate along the computation.
The ``overall error'' describing the gap 
between the intended state of the computer and its noisy state   
takes into account also the cumulated effect of errors from 
earlier computer cycles.

The basic picture we have of a noisy computer is that 
at any time during the computation 
we can approximate 
the state of each qubit only up to some small error term 
$\epsilon$. 
Nevertheless, under the assumptions concerning the errors 
mentioned above, computation is possible. The noisy physical qubits
allow the introduction of logical ``protected'' qubits that are 
essentially noiseless.

What does the error rate $\epsilon$ refer to? 
Perhaps the simplest way to think about it
is as follows. If we measure a qubit (with respect to every basis of 
its Hilbert space) the outcome will agree with the same measurement for the 
intended state, with probability of at least $1-\epsilon$. 
More formally, recall that the trace distance $D(\sigma,\rho)$ 
between two density 
matrices $\rho$ and $\sigma$ is equal to the 
maximum difference in the results of measuring $\rho$ and $\sigma$ in 
the same basis. $~D(\sigma, \rho)=1/2 \|\sigma-\rho\|_{tr}$.  
When the error is represented by a quantum operation $E$ 
the rate of error for an individual qubit is the maximum 
over all possible states $\rho$ of the qubit of the trace 
distance between $\rho$ and $E(\rho)$.

For most of the paper we will consider the same model of quantum computers 
with more general notions of errors. We will study more general 
models for the fresh errors. 
(We will not distinguish between the different components of fresh errors, 
gate errors and storage errors.)
Our models 
require that the storage errors 
not be statistically independent (on the contrary, they should 
be very dependent) 
or that the gate errors not be 
restricted to the qubits involved in the gates and  
be of sufficiently general form. (Note that the errors
may also reflect the translation from this ideal notion of noisy quantum 
computers to a physical realization.) 

There are several other models of quantum 
computers that are equivalent in terms of their 
computational power to the one described here. 
This equivalence does not extend automatically to noisy versions and 
exploring fault tolerance in noisy versions of 
these models is an important challenge in FTQC.  

The discrete models of noisy quantum 
computers we discuss here are direct analogs to continuous-time models 
described, for example, via Lindblad's equations. 
There were some concerns raised by 
quantum computer skeptics that the 
crux of the matter is in the translation from continuous-time models 
to discrete-time models. 
(Those are referred to respectively as Hamiltonian 
modeling and phenomenological modeling 
in Alicki's chapter \cite {Alicki}.) Namely, the concerns 
were that certain Hamiltonian 
models will lead to non-local fresh errors when translated to the 
discrete-time description. Such a possibility was first 
raised and studied by 
Alicki, Horodecki, Horodecki, and 
Horodecki \cite {AHHH}. 
Later, it was suggested 
that non-local 
behavior for the fresh errors can result from ``slow gates'' (see 
\cite {ALZ}), from ``high frequency noise'' (in the Hamiltonian 
model from \cite {TB}), and from ``non-exponential tail'' (\cite {Alicki}).

Recent extensions of 
the threshold theorem to cases that allow time- and 
space-dependence \cite {TB,AGP,AKP,NP} 
start with continuous-time models.

Damaging
models of noise of the kind we describe in this paper can be described 
also for the continuous-time case. 
(Under such models, continuity properties needed to move from continuous-time 
to discrete-time may also pose additional difficulties.)


\subsection {The threshold theorem}

\begin {quote} {\it The existence of fault-tolerant schemes turns the 
problem of building a quantum computer into a hard but 
possible-in-principle engineering problem: if we just manage to store 
our qubits and operate upon them in a level of noise below the 
fault-tolerance threshold, then we can perform arbitrary long quantum 
computations.} --- Kempe, Regev, Unger, and Wolf, 2008 \cite {KRUW}.
\end {quote}
\bigskip

Let $\cal D$ be the following envelope 
of noise operations for the fresh errors: 
the envelope for storage errors ${\cal D}_s$ will consist of 
quantum operations that have a tensor product 
structure over the individual qubits.
The envelope for gate errors ${\cal D}_g$ will consist of quantum 
operations that have a tensor product 
structure over all the gates 
involved in a single computer cycle (more precisely, 
over the Hilbert spaces representing the qubits in the gates). 
For a specific gate the noise can be an arbitrary quantum operation on 
the space representing the 
qubits involved in the gate. (The threshold theorem concerns 
a specific universal set of gates $\cal G$ that
is different in different versions of the theorem.)

\begin {theo} [Threshold theorem] \cite{AB2,Kit1,KLZ}
Consider quantum circuits with a universal set of gates $\cal G$. 
A noisy quantum circuit with a set of gates $\cal G$ and noise envelopes 
${\cal D}_s$ and ${\cal D}_g$ is capable of effectively 
simulating an arbitrary noiseless quantum circuit, 
provided that the error rate for every 
computer cycle is below a certain threshold $\eta>0$. 
\end {theo} 

Here is some further information regarding the threshold theorem:

1. At every computer cycle the specific error operation can be 
chosen from the noise envelope by an adversary. The 
adversary can make his choice based 
on the entire intended evolution and his own earlier choices. 

Part of the fault-tolerance process is identifying 
an ``error syndrome,'' i.e.,   
a set of faulty qubits.\footnote {More precisely, the noise 
is measured in terms
of the tensor product of Pauli operators. The faulty qubits come with 
a Pauli operator indicating the error.}  
Once this is done we can 
give the adversary even greater power to manipulate the faulty qubits 
in an arbitrary way.

2. If we are allowing a smaller numerical value for the threshold $\eta$ 
we can even assume that the envelope for fresh 
errors will be fixed for the entire computation and will 
include at every computer cycle all possible
gate errors (not just gates involved in this computer cycle). 

3. Recent works \cite {TB,AGP,AKP,NP} show that the threshold 
theorem prevails if we allow 
certain space- and time-dependencies for the noise operations. 
For example, 
the quantum computer is described by a lattice in space 
and the fresh-noise envelope allows dependencies among 
qubits that are ``close together.'' A certain amount of dependence of the 
noise on the earlier evolution is also permitted. 

4. The value of the threshold in original proofs of the threshold theorem was 
around $\eta =10^{-6}$ and it has since been improved by at least one 
order of magnitude.  
There are several works showing that under various reasonable assumptions 
on the noise the value of the threshold can be pushed up further. Statistical 
properties 
of the noise, and certain biases, can be used to improve 
the threshold! (See, e.g., \cite {AP}.) 
A breakthrough work by Knill \cite {Kn} uses 
error-detection codes rather than error-correction codes and 
massive post-selection. This allows one to raise the value of $\eta$ (based 
on numerical simulations) to 3\%.  (It also leads to substantially 
higher provable bounds \cite {AGP}.)

5. The threshold theorem relies on a supply of auxiliary 
fresh qubits called ``ancillas.'' Roughly speaking, they are 
needed 
to ``cool'' the system. See 
Section \ref{s:abin} below. 

6. One of the basic properties of FTQC is that the overall error for a single 
physical qubit will be 
bounded above along the entire computation by a small factor times the rate 
of the fresh error. 

7. A weak version of the threshold theorem was first proved by Shor \cite {S3} 
for the case where the error rate is $O(1/\log^c n)$, where $n$ is 
the number of qubits and $c>0$ is some constant. 
Quantum error correction pioneered by Shor himself \cite {S2} and 
by Steane \cite {St} 
plays a crucial role in Shor's as well as all later FTQC schemes.

8. Most proofs of the threshold theorem use concatenation codes. 
A crucial observation that led to an improvement of Shor's result 
was that it is enough to have codes 
that deal well with a random set of faulty qubits.
A  different approach by Kitaev \cite {Kit1,Kit3} is closely 
related to ``topological quantum computing.''
In addition to the reliance on quantum error correction 
the proofs of the threshold theorem also 
rely on a basic theorem of Kitaev and Solovay (\cite {NC}, Appendix 3).

9. The overhead in terms of the number of additional qubits 
needed for fault tolerance is polylogarithmically in the number 
of qubits in the original circuit.

10. FTQC was extended to other 
models for quantum computers. (Let me just mention 
measurement-based models based on cluster states \cite {ND}.) 
The case of adiabatic quantum computation is still open.

The threshold theorem was one of the most outstanding developments 
in the theory of computation towards the end of the last century. It is fair 
to say also that efforts to extend the 
scope of the theorem and to reduce the numerical value of the threshold 
in various situations that can be realistic have demonstrated substantial 
progress over the last decade. As the 
opening quotation of this
section indicates, the threshold theorem may well 
be the basis for the construction of operating quantum 
computers, an achievement that in terms of scientific and technological 
significance 
can be compared to 
the discovery of X-rays 
and their applications towards the first half of the 20th century 
and the construction of digital computers in the second half. 
If quantum computers cannot be constructed then the
detailed understanding of the assumptions in the threshold theorem that fail 
will be a first-rate achievement and 
may also lead to important developments in the theory of 
computation and in physics.

\section {Noisy obstructions} 
\label {s:ob}
\subsection {High-rate noise}

There are several papers showing that if the error rate is 
large then FTQC fails. 
Both in positive and negative results, the 
threshold $\eta$ is not a universal constant but depends 
on the specific assumptions on the noisy quantum computer.
We will restrict our description to 
the case where the computer involves only 1- and 2-qubit gates and 
to depolarizing noise. 

The first negative result of this kind was proved by Aharonov 
and Ben-Or \cite {AB1}. They 
proved that a quantum computer in which every qubit is subject 
to depolarizing noise with probability 97\% in every computer cycle 
can be simulated by a classical computer. 

There are two basic strategies for negative results of this kind.

\begin {itemize}
\item
Strategy 1: After a logarithmic depth computation 
we will not be able to distinguish the noisy output from a random output.
\end {itemize}

Two recent papers in this direction are by Razborov \cite {Ra}, 
who showed that 
FTQC fails when the amount of depolarizing noise exceeds 50\%. 
Kempe, Regev, Unger, and Wolf \cite {KRUW} managed to reduce this bound
to 35.7\%.\footnote{For circuits with unitary $k$-qubit gates, Razborov 
shows an upper bound of $1-\Theta(1/k)$ and Kempe et al. improve 
it to $1-\Theta (1/\sqrt k )$. (Razborov's model is somewhat more general.)}  
Razborov's proof (which follows some ideas from \cite {ABIN} mentioned below) 
is based on tracking the trace distance of the intended state to 
the noisy state of the computer. New ingredients from \cite {KRUW} 
express the effect of depolarizing noise in terms of multi-Pauli operators 
and consider  the {\it Frobenius distance} instead of the trace distance. 

\begin {itemize}
\item
Strategy 2: Efficiently simulate the noisy quantum computation 
by a classical computer.
\end {itemize}

The most recent paper in this direction is by 
Buhrman, Cleve, Laurent, Linden, Schrijver, and 
Unger \cite {BCLLSU}. They showed that a quantum 
circuit cannot be made fault-tolerant against a 
depolarizing noise level of 45.3\%. Their model allows perfect gates from the 
Clifford group and additional noisy one-qubit gates. (For this particular 
model they show that a lower level of noise 
allows universal quantum computing!) While the computational 
complexity conclusion of this strategy is stronger, 
typically it applies to more restricted models (in terms of the set of gates).
 
\subsection {Sequential computation and reversible computation}
\label {s:abin}

We will now describe two additional early negative results 
regarding fault tolerance. 

The first result by Aharonov and Ben-Or \cite {AB1} asserts that 
sequential noisy quantum computers can be simulated by classical computers. 
This result shows that the computational power of decohered 
quantum computers depends strongly on the amount of 
parallelism in the computation. A computation on
the model of noisy quantum circuits with the additional assumption 
that at every round only a single gate is applied can be 
simulated classically. 

The second result is by Aharonov, Ben-Or, Impagliazzo, and Nisan \cite {ABIN}, 
who proved that the computational power of noisy {\bf reversible} 
quantum computers reduces to log-depth quantum computation. 
The proof follows the physics intuition that without 
a cooling mechanism the increase in entropy will eventually 
make computation impossible. A similar result is proved 
for classical computation.\footnote {One of the interesting 
aspects of this paper is a 
beautiful extension to the quantum case of an entropy inequality 
by Shearer \cite {CFGS}.}

\section {Highly correlated noise}
\label {s:hc}

\begin {quote}
{\bf Objection:} {\it Coding does not protect against 
highly correlated errors.}\\
{\bf Response:} {\it Correlated errors can be suppressed 
with suitable machine architecture.} ---John Preskill, 
Quantum Computing: Pro and Con, 1996 \cite {Pre}. 
\end {quote}

\subsection {The error syndrome and error synchronization}
\label {s:es}

The concern regarding highly correlated noise 
has been raised in several papers, yet 
there have been only a few systematic attempts to study what kind of 
correlated errors will cause the threshold theorem to fail.\footnote {Of 
course, 
everyone has always known that the threshold theorem will 
fail for some noise models, e.g., it's hard to protect your quantum 
computer (or digital computer for that matter) 
from a meteor strike. But such models were considered as uninteresting 
and unrealistic.}

Error synchronization refers to a situation where,
while the expected number of qubit errors is small, there is 
a substantial probability of errors
affecting a large fraction of qubits.

A simple way to describe error synchronization is via 
the expansion of the quantum operation $E$ in terms of multi-Pauli operators.
A quantum operation $E$ can be expressed as a linear combination 
$$E=\sum v^I P_I,$$ 
where $I$ is a multi-index $i_1,i_2,\dots,i_n$, 
where $i_k \in \{0,1,2,3\}$ for 
every $k$,  $v^I$ is a vector, and $P^I$ is the quantum 
operation that corresponds to 
the tensor product of Pauli operators whose action 
on the individual qubits is described by the multi-index 
$I$. The amount of error on the $k$th qubit is described by
$\sum \{ \|v^I\|_2^2: i_k \ne 0 \}$. For a multi-index $I$ 
define $|I|:=|\{k:i_k \ne 0\}|$. Let 
$$f(s) := \sum \{ \|v^I\|_2^2: |I| = s\}.$$
We regard $\sum_{s=1}^n f(s) s$ as the {\bf expected amount of 
qubit errors}.

Measuring the noise in terms of tensor product of 
Pauli operators is an important ingredient of several fault-tolerance schemes.
Such a measurement leads to a word $w$ of length $n$ 
in the letters $\{I,X,Y,Z\}$, called the {\bf error syndrome}. We will 
define the {\bf coarse error syndrome} as the 
binary word of length $n$ obtained from $w$ by 
replacing $I$ with '0' and the other letters by '1'. 
Given a noise operation $E$, the distribution $\cal E$ of 
the error syndrome is an important feature of the noise.
Given $E$ we will denote by $\cal D$ the probability distribution 
of coarse error syndrome.
$f(s)$ is simply the 
probability of a word drawn according $\cal D$ having $s$ '1's.

Suppose that the 
expected amount of qubit errors is $\alpha n$ 
where $n$ is the number of qubits.

All noise models studied in the 
original papers of the ``threshold theorem,'' 
as well as some extensions that allow time- and 
space-dependencies (e.g., \cite {TB,AGP,AKP}), have the property that 
$f(s)$ decays exponentially (with $n$) for  $s=(\alpha+\epsilon)n$, 
where $\epsilon >0$ is any fixed real number. 
(This is particularly simple
when we consider storage error, which is statistically 
independent over different qubits.)

In contrast, we say that $E$ leads 
to {\bf error synchronization} if $f(\ge s)$ is substantial 
for some $s \gg \alpha n$. 
We say that 
$E$ leads to a 
{\it very strong} 
error synchronization if  $f(\ge s)$ is substantial for 
$s=3/4-\delta$ where $\delta = o(1)$ as $n$ tends to infinity.
By ``substantial'' we mean larger than some 
absolute constant times $\alpha/s$, or, in other words, 
the multi-Pauli terms for $|I| \ge s$ contributes a constant 
fraction of the expected amount of qubit errors.

\subsection {Examples and models}
\label {s:em}

\begin {prop} 
\label {p:random}
Conditioning on the expected number $\alpha n$ of qubit errors, 
a random unitary operator acting on all the qubits 
of the computer 
yields a very strong 
error synchronization. 
\end {prop}

The proposition extends to the case where we allow 
additional qubits representing the environment.

The proof of Proposition \ref {p:random} is based on a 
standard ``concentration of measure'' argument (see, e.g., \cite {Led}). 
(We will give only  a 
rough sketch.) When we consider 
a typical expression of the form $\sum a_IP^I$ where $\sum a_I^2=1$ 
and $\sum \{a_I^2|I|\}=an$, it will have a large support on $a_{\bf 0}$ 
and the other coefficients will be supported on $a_I$ where $I$ itself 
is typical, i.e., $I$ (the error syndrome) behaves like a random string of 
length $n$ with entries I,X,Y,Z. Hence $|I|$ is around $(3/4)n$.

How relevant is Proposition \ref {p:random}? It is 
well known that random unitary operations on 
the entire $2^n$-dimensional vector space 
describing the state of the computer are not ``realistic'' 
(in other words, not ``physical'' or not ``local''). The best formal
explanation why random unitary operators are ``not physical'' 
is actually computational and relies on the following lemma.

\begin {lemma}
For large $n$, it is impossible to express or even to approximate 
a random  unitary operator using a polynomial-size quantum 
circuit with gates of bounded fanning (namely, gates that operate on a 
bounded number of qubits).
\end {lemma}

An interesting 
problem (posed in \cite {K1}) is to what extent we can describe the 
basic statistical properties of a random unitary operation $U$, conditioned 
on the value of $a(U)$, as the outcome of simple polynomial-size 
quantum circuits. 
As it turns out, 
there are various other 
reasons arising from quantum algorithms to seek computationally 
feasible unitary operators that resemble 
the behavior of random unitary operators. 

Klesse and Frank \cite {KF} described 
a physical system in which 
qubits (spins) are coupled to a bath of massless bosons and 
they reached (after certain simplifications) a noise model with 
error synchronization.\footnote{On the other hand, Shabani \cite {Sha} 
argues that in certain cases correlated errors can lead to better 
performance of quantum codes.}

\subsection {The boundary of the threshold theorem}

For a quantum operation $E$ describing the noise for a quantum computer 
with $n$ qubits we denote by $\alpha(E)$ the expected number of qubit 
errors in terms of the multi-Pauli expansion as described above.

\begin {prop} 
\label {p:o1}
For the known noise models (e.g.,\cite {TB,AGP,AKP})
that allow FTQC via the threshold theorem:

1) The fresh noise $E$ expanded in terms of multi-Pauli 
operations decays exponentially above $\alpha(E)$.

2) The overall (cumulated) noise $E'$ expanded in terms of multi-Pauli 
operations decays exponentially above $\alpha(E')$.
\end {prop}

There is an even simpler property of fresh and cumulated noise for 
noise models for which the threshold theorem holds.

\begin {prop}
\label {p:o2}

For the known noise models (e.g.,\cite {TB,AGP,AKP})
that allow FTQC via the threshold theorem:

3) The fresh noise (at every computer cycle) for almost every pair of 
qubits in the computer is almost statistically 
independent for the two qubits in the pair.

4) The overall noise for almost every pair of qubits in the computer 
is almost statistically independent for the two qubits in the pair.
\end {prop}

Here when we talk about ``almost every pair'' we refer to 
$(1-o(1)){{n} \choose {2}}$ of the pairs when $n$ is large. 

The error syndrome will 
provide a simple way to express correlation between the 
noise acting on two qubits. For two qubits $i$ and $j$, denote by 
$cor_{ij}(E)$ the correlation between the events that the 
qubit 'i' is faulty and the event that the qubit 'j' is faulty. In 
other words, $cor_{ij}(E)$ is the correlation 
between the events that $w_i$ is not $I$, and $w_j$ is not $I$ when $w$ 
is a word drawn according to the distribution of 
error syndromes described by $E$. Proposition \ref {p:o2} implies, 
in particular, that for models allowing the threshold theorem, 
$cor_{ij}(E)$ and $cor_{ij}(E')$ 
are close to 0 for most pairs $i,j$ of qubits.      
(Another simple way to formulate approximate 
independence is in terms of the trace distance 
between the noise operation restricted to two 
qubits from a tensor product operation on these two qubits.) We will 
further discuss two-qubit behavior in the next section.

Note that properties 1 and 3 refer to the noise model, which is one of the 
assumptions for the threshold theorem, while properties 2 and 4 are 
consequences of the threshold theorem and, in particular, 
of suppressing error propagation. For the very basic noise models 
where the storage errors are statistically independent property 3 follows 
from the fact that the number of pairs of interacting qubits at 
each computer cycle is at most linear in $n$. Property 3 continues 
to hold for 
models that allow decay of correlations between qubit errors that depend
on the (geometric) distance between them. Property 1 is a simple consequence
of the independence (or locality) assumptions on the noise for noise models 
that allow the threshold theorem.

\subsection {The rate of highly correlated noise}
\label {s:rate1}

Highly correlated errors are bad for quantum error correction, 
but a potentially  
even more 
damaging property we face 
for highly correlated noise 
is 
that the 
notion of 
``rate of noise for individual qubits''
becomes sharply 
different from the rate of noise as measured by trace distance for the 
entire Hilbert space describing the state of the computer. 

Consider two extreme scenarios. In the first scenario, 
for a time interval of length $t$ there is a depolarizing storage 
noise that hits every qubit with probability $pt$. In the second 
scenario the noise is highly correlated: all qubits are hit with probability 
$pt$ and with probability $(1-pt)$ nothing happens. In terms of the expected 
number of qubit errors both these noises represent the same rate. The 
probability of every qubit being corrupted at a time 
interval of length $t$ is $pt$.
However, in terms of trace distance (and here we must assume 
that $t$ is very small), the rate of the correlated noise 
is $n^{-1}$ times that of the uncorrelated noise. What should be the 
correct assumption for the rate of noise when we move away from the 
statistical independence assumption?

Consider now our first example where 
for an intended state $\rho=U \rho_0$ the noise is described by $U E_0 U^{-1}$.
Since conjugation by a unitary operator preserves trace distance,
the rate of noise in terms of trace distance will not depend on $U$. 
However, the rate of noise in terms of the expected number of qubit errors 
can be much greater. If the unitary operation $U$ that describes 
the computation is ``complicated enough'' that $U E_0 U^{-1}$ 
is highly synchronized, we can even 
expect a situation where the number of qubit errors 
increases linearly with the number of qubits.\footnote {In 
Section \ref {s:dn}  
we will propose (and define formally) ``highly entangled'' 
states as those states 
that are necessarily ``complicated enough.''}

\section {Noise propagation} 
\label{s:np}

\subsection {Noise propagation as a role model}

The basic insight of fault-tolerant quantum computing is that if the
incremental errors are standard and sufficiently small then we can
make sure that the accumulated errors are too.\footnote{By ``standard'' we 
refer to the assumptions that qubits errors are independent 
and that gate errors are confined to the Hilbert space describing 
the qubits of the gates and are independent for different gates. As we 
already mentioned these assumptions can be widened and we can regard 
as ``standard''  
those operations satisfying properties 1 and 3 in 
Propositions \ref {p:o1} and \ref {p:o2}.}
 
 The main issue is 
therefore to understand and describe the fresh (or infinitesimal) noise 
operations. The adversarial models we consider here
should be regarded as models for fresh noise. But the behavior 
of accumulative errors in quantum circuits that 
allow error propagation is sort of a ``role model'' for 
our models of fresh noise.

The common picture of FTQC asserts:

\begin {itemize}
\item
Fault tolerance will work if we are able to reduce 
the fresh gate/qubit errors 
to below a certain threshold. In this case 
error 
propagation 
will be suppressed.
\end {itemize}
 
What we propose is: 
 
\begin {itemize}
\item
Fault tolerance will not work because the overall error
 will 
behave like accumulated errors for 
standard error propagation (for circuits that allow error propagation), 
although {\bf not necessarily} because of error propagation.  

Therefore, for an appropriate 
modeling of noisy quantum computers the 
fresh errors 
should behave like accumulated errors for 
standard error propagation (for circuits that allow error propagation).

(As a result, 
in the end we 
will not be able to avoid error propagation.)\footnote {On the face 
of it, this alternative description looks less natural than the 
common one. The main reason to examine it is in view of 
the extraordinary consequences of the common description.}    
\end {itemize}

Suppose that in your quantum computer at some period along the 
computation, you have two qubits (say, two photons) 
that are entangled. (This entanglement was created along the 
computation and we expect further changes 
in the joint state of these two qubits.) The entanglement between 
the two qubits 
is the result of a chain of gates acting on the computer's qubits and if 
error propagation cannot be suppressed we can expect that the 
accumulated errors for these two qubits will be correlated. But 
there are other reasons for correlation between the errors. The device 
may lead to such a correlation in order to make 
future interaction between the qubits possible. Even if the device 
does not induce such a correlation but pairs of qubits 
are postselected according to the interaction, 
such a postselection may induce correlation 
between the errors. 

The conjectures of this paper amount to saying 
that noise propagation is the 
fundamental property of noisy quantum systems and that we 
need to identify the basic mathematical 
properties of noise propagation and use them in modeling 
noisy quantum computers or noisy quantum 
systems.

\subsection {Forcing noise propagation}

A way to force noise propagation into the model 
is as follows.
Let $K$ is a positive continuous function on [0,1]. We write 
$\bar K(t) = \int_0^tK(s)ds$ and assume $\bar K(1)=1$.
Start with an ideal quantum 
evolution $\rho_t: 0 \le t \le 1$ and suppose 
that $U_{s,t}$ denotes the unitary operator describing 
the transformation from time $s$ to time $t$, ($s<t$). 
Now consider a noisy version with $E_t$ be a 
noise 
operation   
describing the infinitesimal noise at time $t$.
Now replace $E_t$ by 
\begin {equation}
\label {e:master}
E'_t = (1/\bar K(t)) \cdot \int_0^t K(t-s) U_{s,t} E_s U^{-1}_{s,t}ds.
\end {equation}

Relation (\ref {e:master}) represents some sort of smoothing  
of the noise operator in time. If $E_t$ represents standard (local) 
noise operations for noisy quantum computers 
then $E'_t$ will be similar, to some extent, to 
our example from Section \ref {s:ex}. (See below for a discrete-time 
version of equation (\ref {e:master}).)

\begin {quote}
{\bf Main Conjecture:} 
Relation (\ref {e:master}) properly models 
natural noisy quantum systems, and will not allow quantum fault tolerance. 
\end {quote}

For the rest of the paper we will restrict somewhat the class of 
noise operators and 
we will suppose that $E_t$ and hence $E'_t$ are 
described by POVM-measurements (see \cite {NC}, Chapter 2).  

\begin {quote}
{\bf Definition:} Detrimental noise 
refers to noise (described by a POVM-measurement) 
that can be described by equation 
(\ref {e:master}).
\end {quote}

What could be a  motivation for our 
main conjecture? we will mention four reasons:

1) For modeling systems we encounter in nature there is no noticeable 
difference between relation (\ref {e:master}) and the standard 
description of noisy quantum evolution.

2) Regardless of the feasibility of quantum computers, 
noise propagation appears to be the rule for 
open quantum system in nature. Therefore, 
relation (\ref{e:master}) should allow modeling 
information leaks for quantum systems in nature.

3) If FTQC is not possible by whatever fundamental 
principle, 
the conclusion is that noise propagation cannot 
be avoided. 
If noise propagation is a consequence of any hypothetical fundamental 
principle that would cause FTQC to fail, we may as well 
consider noise propagation as such a fundamental principle. 

4) It is expected that the main conjecture will have interesting mathematical 
consequences leading to a coherent picture.

Let me elaborate on the first point. 
For modeling systems encountered in nature the 
standard noise models suffice in the following sense:  
probing the noise in short time intervals is difficult 
and the outcomes may be ambiguous.\footnote{Knowing the 
intended state and the 
noisy state is not enough to determine 
the noise operation uniquely. In addition, we also lose 
information by measuring 
the noisy state.} For longer 
time periods, standard noise models are sufficient to describe 
noisy systems that allow noise propagation 
because moving the incremental noise in 
time will have a similar effect to introducing non-standard 
noise of the kinds we propose. 

More formally, 
we can try to approximate the evolution 
described by (\ref {e:master}) by defining

\begin {equation}
\label{e:reverse} 
E''_t = 1/(1-\bar K(t)) \cdot \int_t^1 E_s K(s-t)ds.
\end {equation}

For noisy quantum computers if $E_t$ represents a standard 
noise operators for every $t$ then so does $E''_t$. (But not $E'_t$.)
Moreover, for systems that do not involve fault tolerance a noisy 
system with the standard noise $E''$ will 
give a good approximation to the system described by the 
non-standard noise $E'$.

We can replace relation (\ref{e:master}) by a discrete time 
description. When we consider a quantum computer that runs $T$ 
computer cycles, we start with standard storage noise $E_t$ for the $t$-step.
Then we consider instead the noise operator 

\begin {equation}
\label {e:discrete-master}
E'_t = 1/(\sum_{s=1}^t K(s/T)) \cdot 
\sum_{s=1}^t K((s-t)/T) U_{s,t} E_t  U^{-1}_{s,t},  
\end {equation}
where again $U_{s,t}$ is the intended unitary operation 
between step $s$ and step $t$.


{\bf Remark:} Since we do not witness quantum error correction in 
nature, understanding the behavior of noisy quantum systems 
where noise propagation is ``forced'' can be of interest 
not just in the context of quantum-computer skepticism. Another 
possibility to force noise propagation is to consider the properties 
of random quantum circuits leading to a given state $\rho$.  
It will also be interesting to examine whether the model of 
noisy adiabatic computers (see \cite {CFP}) 
satisfies our main conjecture.

\section {Detrimental noise}
\label {s:dn}
\subsection {Two conjectures}

\begin {quote}
{\it We can fight entanglement with entanglement.} 
--- John Preskill, Reliable Quantum Computers, 1998 \cite{Pre2}.
\end {quote}
\bigskip


\medskip

In this subsection we present qualitative statements of two conjectures
concerning decoherence for quantum computers  
which, if (or when) true, are damaging for quantum error correction 
and fault tolerance.

The first conjecture concerns entangled 
pairs of qubits.

\begin {quote}
{\bf Conjecture A:}
A noisy quantum computer is subject to error with the property that 
information leaks for two substantially
entangled qubits have a substantial positive 
correlation.
\end {quote}

We emphasize that Conjecture A 
refers to part of the overall error affecting a 
noisy quantum computer. 
Other 
forms of errors and, in particular, errors consistent
with current noise models may also be present.

Recall that error synchronization refers to a situation where,
although the error rate is small, there is nevertheless a 
substantial probability that errors will
affect a large fraction of qubits.

\begin {quote}
{\bf Conjecture B:}
In any noisy quantum computer in a highly entangled state there will be a
strong effect of 
error synchronization.
\end {quote}

We should informally explain already at this point why these 
conjectures, if true, are damaging. We start with Conjecture B. 
The states of quantum computers that apply 
error-correcting codes needed for FTQC are highly 
entangled (by any formal definition of ``high entanglement''). 
Conjecture B will imply that at every computer cycle 
there will be a small but substantial probability that the 
number of faulty qubits will be much larger than 
the threshold.\footnote{Here we continue to assume 
that the probability of a qubit being faulty is small 
for every computer cycle.} This is in 
contrast to standard assumptions that the probability of 
the number of faulty qubits being much larger than the threshold decreases 
exponentially with the number of qubits. Having a small but 
substantial probability of a large number of qubits to be faulty 
is enough to fail the quantum error correction codes.

We move now to Conjecture A. Let us first make the assumption that 
individual qubits can be measured without inducing errors on other 
qubits. This is a standard assumption regarding noisy quantum computers. 
When we start from highly entangled states needed for FTQC and 
measure (and look at the results for) all but two qubits, we 
will reach pairs of qubits (whose intended state is pure) with 
almost statistically independent noise, in contrast to Conjecture A.
Under this assumption it is also possible to deduce Conjecture B 
from Conjecture A. Without making such assumptions 
on measurement, Conjecture A as stated above is not damaging, and 
we will need to extend Conjecture A to disjoint blocks of qubits.

\subsection {Two qubits and two blocks of qubits}

In this  subsection we will describe 
a mathematical formulation of Conjectures A and B. 

The first step in this formal definition is to restrict our attention 
to noise described by POVM-measurements. This is a large class of quantum 
operations describing information leaks from the quantum computer 
to the environment.  

Our setting is as follows. Let $\rho $ be the intended (``ideal'') 
state of the computer and consider two qubits $a$ and $b$. 
Consider a POVM-measurement $E$ representing the noise. We
describe correlation between the qubit errors via 
the expansion in tensor products of Pauli operators, or, in other words,  
by the error syndrome. 

Associated to $E$ (see Section \ref {s:es}) is a distribution ${\cal E}(E)$ of 
error syndromes, i.e.,  
words of length $n$ in the alphabet $\{I,X,Y,Z\}$. A coarser 
distribution ${\cal D}(E)$ of binary 
strings of length $n$ is obtained by replacing the 
letter $I$ with '0' and all other letters by '1'.

As a measure of correlation $cor_{i,j}(E)$ between information leaks 
for the $i$th and $j$th qubit
we will simply take the correlation between 
the events $x_i=1$ and $x_j=1$ according to ${\cal D}(E)$. 
When we have two disjoint sets of qubits $X$ and $Y$ we will 
denote by $cor_{X,Y}(E)$ the correlation between the 
distributions ${\cal D}_X$ and ${\cal D}_Y$, namely, the 
correlation between the distributions of coarse error syndromes 
on these two sets of qubits.

We also define $r_i(E)$ as the probability that 
$x_i=1$ according to the distribution $\cal D$. We let $r_X(E)$ be the average
of $r_i(E)$ for $i \in X$. (To start with, assume that $r_i(X)$ is 
small for every $i$.)

Here and below, $S(*)$ is the (von Neumann) entropy function; 
see, e.g., \cite {NC}, Ch. 11. 
For a set $Z$ of qubits and a state $\rho$  
we denote by $\rho|_Z$ the density matrix obtained after 
tracing out the qubits not in $Z$. If $Z$ contains 
only the $i$th qubit, we write $\rho_i$ instead of $\rho|_Z$.

Suppose that $\rho$ is the intended state of the computer, consider 
two disjoint sets of qubits $X$ and $Y$, let $Z=X \cup Y$,  
and suppose that the joint state 
$\rho|_Z$ is pure (for example, this is the case when $Z$ 
is the set of all qubits). The entropy function $S(\rho|_X)$ is 
a standard measure 
of entanglement between the state $\rho$ on X and on Y. (Recall that in 
this case $S(X)=S(Y)$.) In particular, $\rho$ is a tensor product state
iff the restriction of $\rho$ to $X$ is pure, hence $S(\rho|_X)=0$.

Here is the statement of Conjecture A for two qubits whose 
intended state is pure and an extension to two blocks of qubits.    

\begin {quote}
{\bf Conjecture A:} (mathematical formulation) 
 
(1)(For two qubits in intended joint pure state.)
Suppose that the intended state $\rho$ restricted to $Z=\{i,j\}$ is pure.
\begin {equation}
\label {e:a1}
cor_{i,j}(E) \ge  K(r_i(E),r_j(E)) \cdot S(\rho_i).
\end {equation}

\medskip

(2) (For two disjoint blocks of qubits.)
Let $X$ and $Y$ be two disjoint sets of qubits whose intended joint 
state $\rho$ is pure:
\begin {equation}
\label {e:a2}
cor_{X,Y}(E) \ge  K(r_X(E),r_Y(E)) (\min |X|,|Y|)^{-1} S(\rho|_X).
\end {equation}

\end {quote}
\noindent

Here, $K(x,y)$ is a function of $x$ and $y$ so that 
$K(x,y)/ \min(x,y)^2 \gg 1$  
when $x$ and $y$ are 
positive and small. (Note that Conjecture A(1) does not claim anything 
when the two qubits are noiseless.) If $r_i(E)=r_j(E)=\alpha$ for a 
small real number $\alpha$, then the 
conjecture asserts that $cor_{i,j}(E) \gg \alpha^2$, and, as we will 
see later, this is what is needed to derive error synchronization.

The main mathematical challenge 
is to show that Conjecture A is satisfied when we force noise 
propagation, for example, via relation (\ref {e:master}). 

\begin {quote}
{\bf Main mathematical conjecture:} 
The assertions of Conjecture A are satisfied
for noisy quantum computers where the noise is 
described by equation (\ref {e:master}).
\end {quote}

It will be interesting to check whether the assertion of Conjectures A and B 
holds for noisy adiabatic computers and also 
for our very first example from Section \ref {s:ex}.   

We mention a second mathematical conjecture related to Section \ref {s:rate1}.
\begin {quote}
{\bf Second mathematical conjecture:} For noisy quantum computers described 
by relation (\ref {e:master}), the rate of fresh noise in 
terms of the expected number of faulty qubits 
scales up linearly with the number of qubits if the 
intended state is highly entangled.  
\end {quote}

{\bf Remarks} 

1) As an alternative  measure of entanglement we can simply take 
the trace distance between the state induced on the two qubits 
(or, more generally, two disjoint sets of qubits) and a separable state.
Formally, let $SEP(A,B)$ denote the set of mixed 
separable states on $A \cup B$, 
namely, states that are mixtures 
of tensor product pure states $\tau=\tau_A \otimes \tau_B$. Define 
$Ent(\rho:A,B) =max \{ \|\rho_{A,B}-\psi\|: \psi \in SEP(A,B) \}$.

2) We will use only Conjecture A for the cases where the intended 
joint state
is pure. The conjecture itself extends to the case where 
the intended joint state is not pure. If we use the 
trace distance from a separable state as the 
measure for entanglement then the conjecture carries over without change.
If we want to use an entropy-based  measure we can use the 
minimum of the relative entropy $S(\rho|_{X \cup Y}\|\psi)$ 
over all $\psi \in SEP(A,B)$. 

\subsection {Why are the conjectures damaging?}

We already described why error synchronization fails 
current methods for fault tolerance. We need to describe formally 
Conjecture B and explain why Conjecture A implies Conjecture B.

\begin {prop}

\label {p:cor2q}
Let $\eta<1/20$ and $s>4\eta$.
Suppose that $\cal D$ is a distribution of 0-1 strings of length $n$
such that $p_i({\cal D}) \ge \eta$ and $c_{ij}({\cal D}) \ge s$. Then
\begin {equation}
{\bf Prob} (\sum_{i=1}^n x_i > sn/2) > s\eta/4. 
\end {equation}

\end {prop}

The proof of this proposition is indicated in \cite {DD} and we expect 
that a similar argument will also yield:

\begin {prop}

\label{p:corpart}
Let ${\cal D}$ be a probability distribution on 0-1 strings 
of length $n$. Suppose that for a random partition of the bits 
into two parts $X$ and $Y$, 
the expected value of the correlation satisfies:
$${\bf E}(cor{\cal D}(X,Y)) \ge s.$$ 
Then  
\begin {equation}
{\bf Prob} (\sum_{i=1}^n x_i > sn/2) >  s\eta/4. 
\end {equation}

\end {prop}


We can now state formally also Conjecture B. 
The notion of ``highly entangled state'' in Conjecture B can be 
taken as a state for which when we partition the qubits into 
two parts at random the expected amount of entanglement between the 
two parts is large. This is indeed the case for states used for 
error correction. With this definition, Proposition \ref {p:corpart} asserts 
that Conjecture A(2) (for disjoint blocks of qubits) 
implies Conjecture B.

{\bf Remark:}  The following critique of the possibility of any systematic 
damaging relation between the state 
of the quantum computer and the noise was raised by several people. 
Having a classical computer control a quantum computer 
makes it  possible  to run a variant of any quantum computer program where at 
the initial state we apply  random Pauli operators on every qubit 
and modify the action of the gates accordingly. In this way the 
state of the quantum computer will always be the same mixed 
state for the entire computation. A detailed proof of such a result along 
with an interesting interpretation and discussion was offered by 
Dorit Aharonov \cite {Dorit}. (Her work extends and relies on earlier works by 
Preskill, Shor and Ben-Or.) 

A response to this critique is 
based on the following 
point  made by Aharonov in the same paper. Consider the qubits of the 
mixed-state quantum computer, together with the qubits (which are simply 
random bits) of the  
computer that controls its state, as a single larger 
pure-state quantum computer. We assume that the quantum qubits are 
noisy but the control classical bits are noiseless. 
Then (with very high probability) 
there will be a strong entanglement 
when we partition all the qubits into two parts. Conjecture A will imply that 
a large correlation between information leak for the two parts. 
Now we can apply a variant of Proposition \ref {p:corpart} to deduce 
strong error synchronization 
for the noisy quantum qubits
and hence the failure of FTQC.   

We conclude this subsection with a description of another avenue 
(\cite {K2,DD}), which goes from the two-qubit case of 
Conjecture A (extended in another direction) to 
Conjecture B. This goes through a notion 
of ``emergent entanglement.''  
The emergent entanglement of two qubits is the maximum expected amount 
of entanglement between two qubits 
when the other qubits are measured (separably) and we look at 
the outcome of the measurements.
(This is a less drastic notion than the definition in \cite {DD,K2}, 
which appears to be too strong.) 

In standard noise models for quantum computers, measuring 
and looking at the results for all but two qubits of the computer will not 
affect the errors on these two qubits.  
We can define a highly 
entangled state as a state where the expected emergent 
entanglement among pairs is large. This is the case for states used 
in quantum error correction. A strong form of 
Conjecture A is obtained if we take emergent entanglement as 
the measure of entanglement. Using Proposition \ref{p:cor2q}, this 
strong form of Conjecture A for pairs of 
qubits implies Conjecture B.

\subsection {Censorship}

The conjectures regarding noisy quantum computers 
and error synchronization are rather counterintuitive. The possibility that 
when the state of the quantum computer is highly entangled 
then for the period of time when the probability of every 
qubit being corrupted is very small there will still 
be a substantial probability of a large fraction of faulty qubits seems  
strange. One comment is that the argument will be to some extent 
counterfactual and that these properties of noise will 
imply severe restrictions on feasible states of noisy 
quantum computers. The counterintuitive forms 
of noise will occur 
for infeasible states. (Yet the conjectures on the 
nature of noise can be tested on feasible states.)

Computational complexity poses severe restrictions on the feasible 
states of (noiseless) quantum computers. For 
example, as we already mentioned, a state that
is approximately the outcome of a random unitary operator 
on the entire $2^n$-dimensional Hilbert space is computationally out of 
reach  when the number of qubits is 
large.

Adversarial forms of noise may lead to further 
restrictions on feasible states for noisy quantum computers. 
Here is a specific conjecture in this direction (partially 
responding to a challenge posed by Aaronson in \cite {Aa1}.) 
We assume that the ``ideal'' state of the quantum 
computer (before the noise is applied)
is a pure state. (Some adjustment to our conjecture 
will be required when the 
ideal state itself is a mixed state.) 

Let $\rho$ be a pure state on a set 
$A = \{ a_1,a_2,\dots,a_n\} $ of $n$ qubits. Define

$$ENT(\rho; A) = -S(\rho)+ \max S(\rho^*),$$ 
where $\rho^*$ is a mixed state with the same marginals
on proper sets of qubits as $\rho$, i.e., 
$\rho^*|_B = \rho|_B$ for every proper subset $B$ of $A$.

Next, define

$$\widetilde{ENT}(\rho) = \sum_B \{ENT (\rho; B): B \subset A \}.$$

\noindent
In this language a way to formulate the censorship conjecture is:

\begin {quote}

\noindent
{\bf Conjecture C}: 
There is a 
polynomial $P$ (perhaps even a quadratic polynomial) such that 
for any quantum computer on $n$ qubits, 
which describes 
a pure state $\rho$, 
\begin {equation}
\label {e:c}
\widetilde{ENT} (\rho) \le P(n). 
\end {equation}
\end {quote}

The parameter $\widetilde{ENT}$ can serve as an alternative
measure for the notion 
of ``highly entangled states'' from Conjecture B. 
States (admitting some symmetry 
in order to ensure that the entanglement is not 
confined to a small subset of qubits) 
where $\widetilde{ENT}$ is 
quadratic (perhaps even super-linear) in the number of qubits can already be 
regarded as ``highly entangled.''

\subsection {Testing it}
\label {s:ti}

\begin {quote}
{\bf Objection:} {\it In the near term, experiments with quantum 
computers will be mere demonstrations. They will not teach us anything.}\\
{\bf Response:} {\it ...We will learn about 
correlated decoherence.}--- John Preskill, Quantum Computing: 
Pro and Con, 1996 \cite {Pre}.
\end {quote}
\bigskip

The 
conjectures regarding 
pairs of qubits or error synchronization can 
be examined on rather 
small quantum computers.

For example, under the standard assumptions on noise, 
a circuit able to correct 
two errors will be able to create pairs of entangled 
qubits with almost independent errors even if gates used 
in the circuit each have a small but otherwise arbitrary form of errors.
(This will require a small overhead on the rate of error.) 

Creating pairs of entangled qubits (say, EPR pairs) 
with almost uncorrelated errors, which 
runs counter to our conjectures, can be tested on a rather small 
quantum computer with 10-20 qubits.

Here we propose to test 
properties of the overall (cumulative) noise. It is probably harder to 
probe the ``fresh noise'' directly (and the outcomes will be 
less conclusive), but probing ``fresh noise'' will enable one to  test these 
ideas 
for systems operating already with a small number of qubits. 
Some detrimental noise behavior may be witnessed in the realization of 
quantum error correction for a single error. 

An important experimental quantum error correction 
challenge is the ability 
to approximate in small quantum computers 
every possible pure state on a few qubits (three, four, five). Achieving
this will go a long way toward refuting the conjectures on detrimental noise.

One point to notice is that the conjectures we 
consider in this paper are not equivalent to the familiar concerns 
about scalability of quantum computers. Our conjectures 
may come into play, as anticipated 
in Preskill's quotation starting this section, already with small quantum 
computers.

Two warnings are in order:

1) Empirical support for the conjectures from one device 
does not apply to other devices. The mechanism leading to the conjectured 
behavior is not universal but may depend on the device.

2) We need a low error rate to start with. In order to 
identify the effects of non-standard noise we still need 
to suppress standard noise of a higher rate.

Here is an example. Most current implementations of ion trap computers 
creating entanglement between two qubits require physically 
moving them together. This suggests that for these ion trap computers
fresh errors will be 
correlated for {\bf every} pair of qubits and that using 
them to create entangled pairs of qubits with 
uncorrelated errors will not be possible. Of course, this suggestion
should be tested experimentally. (While it seems rather clear that 
for these ion trap computers, fresh errors for every 
pair of qubits are going to be correlated, the stronger claim of 
positive correlation for information leaks is not clear.)

In principle, for some other implementation of ion trap computers it may be 
possible to induce entanglement between pairs of qubits 
without affecting any other qubits.

\section {Detrimental noise for general quantum systems} 
\label{s:gs}

{\it Right now the only way I can see engineering 
worlds with classical but not quantum computation is to 
engineer a world in which  ``phase''-type decoherence is massive or 
crazily correlated but ``amplitude''-type decoherence 
is not.}---Dave Bacon, The Quantum Pontiff, 2006.

\subsection {Our first example revisited}

Consider our first example of a quantum computer where 
when the 
quantum memory is in a state $\rho$ and $\rho= U \rho_0$, the noise
$E$ will be $U E_0 U^{-1}$.
When we try to describe the relation between 
the state of the computer and the noise, this example describes, 
for every state
$\rho$, an envelope of noise $D_\rho = \{U E_0 U^{-1}: U \rho_0 = \rho\}$.
This is a huge class of quantum operations most of which are 
irrelevant (being computationally infeasible.) An important  
property of this noise is: 

\begin {equation}
\label {e:main}
{\cal D}_{U\rho} = U {\cal D}_\rho U^{-1}.
\end {equation}

Relation (\ref {e:main}) amounts to saying that 
there is a component of quantum noise that is invariant under unitary 
operations and thus does not depend on the device that 
carries these operations.

{\bf Remark:} 
Note that contrary to Bacon's assertion quoted 
at the beginning of this section, 
our conjectures on the nature of noise do not treat 
amplitude errors and phase errors differently. Rather, the conjectures and 
especially relation (\ref {e:main}) do precisely the opposite
in asserting that some ingredient of noise 
is inherently invariant under symmetries of the Hilbert space 
describing the states of the computer. 
Such a symmetry for decoherence may account 
for the symmetry-breaking leading to 
the classical behavior of large quantum systems.

\subsection {Noisy quantum systems}

When we talk about general noisy quantum systems and not about 
controlled systems 
with a clear ``intended'' evolution there is no 
obvious meaning to the notion of ``errors.'' There are two 
related issues to consider:

\begin {enumerate}
\item
Information leaks from the system to its environment. 
\item
Errors in any {\bf description} of the evolution of a noisy quantum system.
\end {enumerate}

As before, we restrict our attention to noise described by 
POVM-measurements.

We can now ask: what are the laws of decoherence 
for general noisy quantum systems 
that follow the properties of noise propagation?

As with the case of standard models of noise, we would like to describe an 
envelope of noise, i.e., a large set of quantum operations, so that 
when we model noisy quantum operations or more general processes 
the incremental (or infinitesimal) noise should be taken from this envelope. 
Conjectures A and B and our first example propose some systematic 
connection between the noise and the state. However, in these conjectures
both the assumption in terms of entanglement and the conclusion in terms of 
correlation rely on the tensor product structure of $\cal H$. 

Here is a (rather tentative) proposal on how to 
formalize this connection for general systems:

\begin {quote}
{\bf Definition:} A 
D-noise of a quantum 
system at a state $\rho$ is a quantum 
operation $E$ that commutes
with 
some non-identity unitary quantum operation that stabilizes $\rho$.
\end {quote}

This definition describes a (huge) class ${\cal D}_\rho$ of quantum 
operations that respect the relation 
${\cal D}_{U\rho} = U {\cal D}_\rho U^{-1}.$

\begin {quote}

{\bf Conjecture D:} D-noise cannot be
avoided in every noisy quantum process. 

\end{quote}

On its own our suggested definition of D-noise is extremely 
inclusive, and so is any (nonempty) envelope of noise operations 
that satisfies relation (\ref {e:main}). For 
example, a D-noise on a state of the form $\rho \otimes \rho$ can be 
standard even if $\rho$ is highly entangled.   
However, there are two additional conditions we have to keep in mind:

\begin {enumerate}
\item
The hypothesis that the overall noise contains a large 
D-component applies to every subsystem of our original system. (An appropriate
``hereditary'' version of Conjecture D may suffice to imply Conjectures A 
and B for noisy quantum computers. This has yet to be explored.)
\item
The operation describing the noise should be ``local'' or more 
formally (see Section \ref {s:em}) 
``computationally feasible'' 
in terms of local operations describing the system.
\end {enumerate}

Trying to express the noise envelope  
in terms of the entire evolution (not just the temporal state) or 
in terms of a set of ``gates'' that describe the evolution may lead 
to sharper descriptions. We will not pursue these directions here. 
Another interesting issue is extending the censorship 
conjecture (Conjecture C)  
to general quantum systems. This conjecture and the whole 
notion of entanglement rely on a tensor product structure that we do 
not have in the general case. It is true in rather general cases that a tensor
product structure emerges (not necessarily the ``natural'' tensor product 
structure). It is not known how general this 
phenomenon is. And we can ask whether Conjecture C would extend 
to arbitrary noisy quantum systems for some emerging tensor power structure.

\section {Linearity, causality, memory, and rate}
\label {s:cr}

\subsection {Some concerns}
\subsubsection {Linearity}

Our conjectures for noisy quantum computers and for noisy quantum systems
amount to nonlinear relation between the noise envelope and the 
state of the computer (system). Such nonlinear relations 
do not violate linearity of quantum mechanics. For example, 
if we consider the noise in our main relation (\ref {e:master}) or 
in our opening example 
as a function of the entire earlier evolution then it is completely linear.  
Nonlinearity is caused by ignoring the earlier evolution and considering 
the relation between the noise and the state for all possible 
evolutions leading to this state.

\subsubsection {Memory}


Do our conjectures and relation (\ref {e:master}) mean 
that the environment necessarily ``memorizes'' 
the past evolution, or, at least, a very crude property 
of the past evolution encoded by the 
noise envelope? In order to relate to this question we note 
that while the models of noisy quantum 
evolutions and noisy quantum computers are sufficiently rich to 
model any noisy quantum evolutions that we can imagine or create, 
these models can give wrong or incomplete intuitions regarding issues 
like memory and causality. The distinction between the quantum 
computer that performs the intended evolution and the environment 
that induces the noise is a property of the mathematical model and not 
a description of the physical reality. The mathematical dependence of the 
noise on the past evolution can represent the effect of the past evolution 
on the environment, but it can also represent various other things, such 
as consequences of the feasibility of the past evolution 
on the physical device performing it, and postselection. 

\subsubsection  {Causality}

\begin {quote}
{\it When a gun shows up in the first act, it will go off in 
the third.} Chekhov's gun principle.
\end {quote}

Consider an intended pure-state evolution $\rho_t$, $0\le t\le 1$ of 
a quantum computer, and a noisy realization $\sigma_t$, $0 \le t \le 1$. 
Assuming that $\sigma$ is 
close\footnote {In some sense, e.g., in terms of the 
expected number of qubit errors.} to $\rho$ for the {\bf entire} 
time interval may create a systematic relation 
of the infinitesimal noise at an 
intermediate time $t$ on the entire intended evolution of $\rho$.       

It is a {\bf consequence} of FTQC that dependence of the errors on the 
past evolution and on the future (intended) evolution becomes 
negligible. 

Of course, we need {\bf also} to be able to describe the noise as the 
outcome of a local, computationally feasible 
process which depends only on the past. (Indeed 
relation (\ref {e:master}) offers such a description.) 

Perhaps the following (completely classical) example can shed some 
light on the 
causality issues we discuss. Suppose that an airport-averse professor 
is planning a trip as follows: 
Leave Davis at 8:00; arrive at San Francisco 
airport at 9:30; take the 10:00 flight to Chicago; present a 
lecture the next morning at UC at 11:00. 
Assuming that this plan 
is realized up to small errors we can deduce that it is much more likely that 
the professor arrived at the airport earlier rather than later. 
The errors compared to the original plan may thus depend 
on the entire planned evolution (assuming its success). 

Of course, it also necessary that the errors compared to
the planned
time-estimate to reach the airport 
can  be described as consequences 
of events 
occurring 
in California 
before arriving at the airport, e.g., the number of 
people taking the highway being less than 
average.\footnote{This can also be influenced by a future event, e.g.,    
 a major sports game shown on TV that evening at 10.}

\subsubsection {Faraway photons}

Suppose we have two faraway photons at a given entangled state at time $T$.
Consider their state at time $T+t$. Is there any reason to believe 
that the changes will not be independent? We can expect detrimental noise 
at the time the entanglement is created but we cannot expect 
it at a later time. Is this a counterexample to our conjecture 
regarding pairs of qubits? 

We relate to this concern in the next subsection.

\subsection {Modeling the rate of noise for noisy quantum evolutions}

\begin {quote}
{\it The physical systems in which qubits may be 
implemented are typically tiny and 
fragile (electrons, photons, and the like). This raises the 
following paradox: On the one hand we want to isolate these systems 
from their environment as much as possible, in order to avoid 
the noise caused by unwanted interaction with 
the environment --- so called ``decoherence.'' But on the other 
hand we need to manipulate these qubits very precisely 
in order to carry out computational operations. A certain level 
of noise and errors from the environment is therefore 
unavoidable in any implementation.} --- Kempe, Regev, Unger, 
and Wolf, 2008 \cite {KRUW}.  
\end {quote}
\bigskip

The quote from Kempe et al. 
points to some genuine 
difficulty in modeling noisy quantum systems. We can exhibit 
extremely stable entangled quantum states, and yet we believe that 
quantum systems are inherently noisy. We can also have isolated 
qubits that do not interact at all that are subject 
to uncorrelated noise, and yet we propose in this paper 
that for the appropriate model 
of noisy quantum computers the noise should be highly correlated. 
The noise (its rate and its form) depends on the 
fact that we need to manipulate the qubits, 
but what is the formal description of such a dependence? 

When we model the fresh (or infinitesimal) noise for the 
evolution of a noisy quantum computer or even a 
general noisy system, what should be a lower bound on 
the rate of noise? This is an interesting 
issue even when it comes to a single
noisy qubit.

Recall that the usual assumption regarding 
the rate of noise is that for every qubit the probability of it being faulty 
is a small constant for every computer cycle. 
We propose the following refinement of this assumption.

\begin {quote}
{\bf Conjecture E:} 
A noisy quantum computer 
is subject to (detrimental) 
 noise with the following property: 
the rate of noise at time $t$ (in terms of trace distance) is 
bounded from below by a measure of 
noncommutativity 
between 
the operators describing the
evolution 
prior to time $t$ and those describing it after time $t$.  
\end {quote}

The lower bound according to Conjecture E for the rate of noise 
when the process starts or 
ends is zero. The rate of noise can also vanish for classical 
systems where all the 
operations commute. Conjecture E can be regarded as a proposed 
refinement on the assumptions regarding the rate of noise 
even for a single qubit.

\section { Discussion}
\label{s:di}

\subsection {Classical noisy systems}


\begin {quote}
{\it When it rains it pours.} English proverb.\footnote {Similar 
proverbs asserting that troubles come together exist in other languages.}
\end {quote}

Our definitions of detrimental noise and our various 
conjectures as stated here 
do not have any implications for classical noisy systems. 
Still, some of our conjectures were 
originally formulated  also for
 ``natural'' noisy classical 
correlated systems; see \cite {K2}. (As a matter of fact, the 
behavior of classical noisy systems was one of the motivations 
for Conjectures A and B.)

For example, we can expect error synchronization for 
attempts to describe (or prescribe) noisy highly correlated 
stochastic systems such as the weather or the stock market.

Understanding noise and the study of de-noising methods span wide areas. 
For example, in machine learning we can see the example 
where text and speech represent respectively the intended (ideal) and 
noisy signals. Certain statistical methods of de-noising
are based on assumptions that run counter to our conjectures. 
However, our conjectures 
are in agreement with insights asserting that such 
statistical de-noising methods will leave
a substantial amount of noise uncorrected. (Moreover, ``natural'' examples of 
noisy highly correlated classical systems exhibit a moderate 
degree of dependence, much 
less than the sort of dependence required 
for quantum error correction and various basic quantum algorithms.)

Because of the heuristic (or subjective) nature of the notion 
of noise in classical systems (and of the notion of probability itself),
such a formulation, while of interest, leads to several 
difficulties. Moreover, we can exhibit counterexamples to 
classical analogs of Conjectures A and B  based on the 
ability to have noiseless classical memory and computation.
Therefore, the analogy 
with a classical noisy system does not make the conjectures of 
this paper more compelling (or less compelling) but rather gives 
a wider context 
in which to discuss them.

Let me mention a question that is often raised in 
discussions on quantum fault tolerance and deserves better 
understanding.

\medskip
\begin {quote}
How is it  possible that quantum fault-tolerant computation fails while 
classical fault tolerant computation succeeds?
\end {quote}
\medskip

One conceptual difference between quantum and classical 
error correction (mentioned in \cite {K2}) is that clean bits can be extracted 
from noisy signals that do not erase all information. (For example, 
when you have a stream of bits and every bit is replaced 
by a random bit with a probability of 0.9999.) However, in the quantum case, 
there is a whole range of noise that does not enable extracting 
clean qubits from a stream of noisy qubits. (Extracting clean {\bf bits} 
is still possible.) Another interesting conceptual difference 
related to error correction of correlated noise (with a single error) is 
described by Ban-Aroya, Landau and Ta-Shma \cite {BLT}. The paper of 
Alicki and Horodecki \cite {AH} can also be regarded 
as a proposed explanation.


\subsection {Theoretical and empirical physics}

{\it The development of the theory of quantum error correction may 
in the long run have broader and deeper implications than the development
of quantum complexity theory} --- John Preskill, Quantum Computing: 
Pro and Con 1998, \cite {Pre}. 

\medskip

Implementing quantum error correction requires complicated 
and specific constructions of quantum processes 
that we do not encounter in nature. There are, however, interesting 
suggestions regarding usefulness of quantum error correction 
in the study of black holes \cite {HP}, quantum gravity, and 
other areas (see \cite {Pre4}). 
On the other hand, there has been little 
effort 
within the QM framework to understand 
what could be 
the implications  
of failure of FTQC for physics. Some proponents of 
quantum computers regard 
the feasibility 
of computationally superior quantum computers  
as a logical consequence of quantum mechanics. 
Some skeptics regard it as a not particularly interesting far-fetched idea.

An obvious concern regarding adversarial noise models 
(and other skeptical claims about quantum computers) 
is whether they are consistent with well-established 
phenomena from physics and current empirical 
evidence. 
For example, are such noise models  
consistent with superconductivity?
Since detrimental noise appears to express familiar 
properties of noise propagation it seems reasonable that detrimental noise 
is consistent with the physics that we see around us, but this deserves 
much closer examination.

On the other hand, detrimental noise is 
in conflict 
with hypothetical physics constructions. The construction 
of stable non-Abelian anyons \cite {Kit3,MR} might be inconsistent 
with the conjectures regarding detrimental noise since (at least according
to some models describing them) such non-Abelian anyons demonstrate 
quantum error correction based on highly entangled systems. 

A potential implication of a deeper understanding of 
quantum error correction (and their limitations) 
may lead to a better understanding of the 
 ``quantum measurement 
problem'' \cite {Leggett}\footnote{Leggett's view in \cite {Leggett} is that 
regarding the  phenomenon of decoherence as an explanation 
of the ``measurement paradox'' is a ``gross logical fallacy.'' Perhaps, 
contrary to his view, the crux of the matter resides in a deeper 
understanding of the phenomenon of decoherence itself.} 
as well as to:

\begin {quote}
{\bf Conjecture F:} Stable non-Abelian anyons do not 
exist in nature and cannot be created.
\end {quote}

\medskip

\begin {quote}
{\bf Third mathematical conjecture:} (i) Show that the model of noisy 
quantum evolutions with forced noise 
propagation (relation (\ref {e:master})) and the conjectures on 
the relation between the noise envelope and the state
do not support non-Abelian anyons.

(ii) Show that this model and these conjectures 
do support Abelian anyons as well as even 
more basic quantum mechanics phenomena.
\end {quote}

\subsection {Classical simulation of noisy quantum systems}

Here is an interesting question: 
\begin {quote}
Does a (hypothetical) failure 
of computationally superior quantum 
computers necessarily mean that classical computers 
are capable, in principle, of simulating efficiently the behavior of
the quantum processes we witness in nature? 
\end {quote}

Of course, we can ask if in view of the complex nature of 
fault tolerance based on quantum error correction 
classical computers are capable, in principle, of simulating 
natural quantum processes, anyway (even if quantum computers are feasible).
Candidates for processes 
that may occur in nature and possibly hard to simulate classically 
are distributions represented by bounded-depth 
quantum circuits (even random such circuits). Understanding 
the computational complexity of such distributions is a 
question of great importance.

\subsection {Engineering, science, and time}

One of the interesting aspects of quantum error correction (and of 
quantum information in general) is the mixture of 
theory and practice, science and engineering, and various areas 
of mathematics, physics, and computer science (and more). It is often 
the case that the borders between engineering issues 
and abstract theoretical and conceptual matters 
are rather blurred. We will mention one example.  

In his paper \cite {Pre} Preskill (see also \cite {Pre4}) proposes small 
quantum computers with quantum error correction 
capability as a way to engineer 
more accurate clocks than those available at present. 
Far-fetching (and flipping)
Preskill's suggestion we can ask: Does a failure of FTQC (in principle) 
have any conceptual bearing on the 
notion of time itself?

\subsection {Computational complexity issues}
\label {s:cs}

The foundations of noisy quantum computational 
complexity were laid by Bernstein and Vazirani in \cite {BV}.
The problem of describing complexity classes of quantum computers
subject to various  models of noise was proposed 
by Peter Shor 
in the nineties. (Although we naturally expect computational 
power between BQP and BPP it is possible, in principle, 
that certain noise models will allow efficient algorithms 
even for problems not in BQP.) 
Scott Aaronson 
\cite {Aa1} 
asked for the computational complexity consequences of various hypothetical  
restrictions on feasible (physical) states for 
quantum computers. In particular, he posed the interesting ``Sure/Shor 
challenge'': to describe such restrictions  
that do not allow for polynomial-time factoring of integers
and at the same time do not violate what can already be 
demonstrated empirically.  

The threshold theorem and some of 
its recent versions give a 
fairly good description of the wide models of noise that 
allow universal quantum computing when the noise rate is sufficiently small.
We mentioned several results (\cite {ABIN,Ra,KRUW}) 
showing that for the standard noise models 
when the computation is reversible or when 
the noise rate is high, the computational power 
reduces to BPP (for some results) or $BPP^{BQNC}$ (the power of classical 
computers together with log-depth quantum circuits). 
(This is sufficient for polynomial-time factoring! 
Cleve and Watrous \cite {CW} gave a polynomial  
algorithm for factoring that requires, beyond classical computation, only 
log-depth quantum computation.) 

How bad can the effect of correlated errors be? I tend to think that 
for an arbitrary form of noise, if the 
expected numbers of qubit errors in a computer cycle is sufficiently 
small then problems in $BPP^{BQNC}$ and, in particular, 
polynomial-time factoring can prevail.
A rough argument in this direction would go as follows. First replace 
a given log-depth circuit by a larger one capable of 
correcting standard errors; then run the computation 
a polynomial or quasi-polynomial (depending on the precise 
overhead in the fault-tolerant circuit) number of times to 
account for highly synchronized errors. 

On the other hand, it may be possible (but not easy) to prove that highly 
correlated errors of the kind under consideration do not allow fault 
tolerance based on quantum 
error correction, and perhaps also that they suffice to 
reduce the computational power to  $BPP^{BQNC}$. 

The most interesting direction, in my opinion, would be 
to show that with the full power of 
detrimental errors, e.g., as defined in equation (\ref {e:master}), 
including the conjectured effect on 
the expected number of qubit errors in one 
computer cycle (Section \ref {s:rate1}), the computational 
power of noisy quantum computers reduces to BPP.

\subsection {An analogy: the $NP \ne P$ problem}

In this  section we draw a quick analogy 
between the $NP \ne P$ problem and a skeptical point of view 
regarding quantum computers. (An analogy in the opposite 
direction, namely, a parallel discussion of reasons 
to believe $NP \ne P$ and the feasibility of quantum computers,  
is offered by Aaronson \cite{A:R2B}.)

A point of view that can be found implicitly or implied 
even today, and certainly could have be found much more 
in the middle of the twentieth century, asserts that: 

\begin {quote}
{\it Every finite problem can be solved, in principle, by a 
digital computer.}\footnote{Of course, the opposite point 
of view was also present since the early days of computers. But it was 
not until the seventies that it was fully realized that the limitation of 
computers is an important scientific question.}
\end {quote}

The major scientific change that happened in the last fifty years is 
twofold. 
%
Attempts to find algorithms for certain ``hard'' problems, or  
special-purpose computing devices to deal with them, have failed.
In addition, the conjecture that some problems are 
computationally infeasible was stated 
in concrete mathematical terms and has led to an elegant, coherent, 
and rich mathematical theory of computational complexity. 

This has led to the modern point of view where the common wisdom is: 

\begin {quote}
{\it Large NP-complete problems cannot be solved by any 
realistic computational device.}
\end {quote}

Of course, when an algorithm or a physical device whose purpose 
is to solve NP-complete problems is offered it is not always easy to explain 
why it is going to fail, and there is no ``universal'' reason or mechanism for 
such a failure. Often, this leads to rather interesting research.

We come to the issue at hand, starting with a common belief or assumption that
 every quantum state that we can imagine can, in principle, 
 actually be created.

It is clear that 
computational complexity does restrict the type of states 
that can be created, and this agrees with earlier physics insights.
So a realistic version is:
Every (computationally feasible) state that we can imagine 
can, in principle,  
be created.
%

The research of 
the past fifteen years has led to a more detailed insight centered 
around quantum error correction.   

\begin {quote}
{\it Every (computationally feasible) state that we can imagine 
can, in principle,  
be created via quantum error correction.}
\end {quote}

The alternative possibility proposed here and in various other papers is:

\begin {quote}
{\it Highly entangled states cannot be created; 
full-fledged quantum error correction 
is not feasible, nor are computationally superior quantum computers.} 
\end {quote}

In order to support such a position we will need  
strong experimental evidence, most 
importantly, strong evidence that attempts to build quantum 
computers face some solid obstacles already for handful of qubits.  
In addition, we need a  
coherent and elegant mathematical explanation for a 
principle that can imply that quantum error correction and quantum computers 
fail. In words, the principle we propose in this paper 
is a familiar one: ``noise propagation.''\footnote{In a few more words: 
``the process for creating entanglement necessarily leads to noise 
synchronization.''} It comes with the 
important asterisk that properties of noise propagation can have other causes. 
A main point of this paper is that 
there is more to be explored and understood 
in the mathematical study of noise propagation.

\end {document}